\documentclass[twocolumn]{aastex63}

\usepackage{textcomp, gensymb, amsmath}

\shorttitle{The Maximum Mass-Loss Efficiency}
\shortauthors{Vissapragada et al.}

\begin{document}

\title{The Maximum Mass-Loss Efficiency for a Photoionization-Driven Isothermal Parker Wind}

\correspondingauthor{Shreyas~Vissapragada}
\email{svissapr@caltech.edu}

\author[0000-0003-2527-1475]{Shreyas~Vissapragada}
\affiliation{Division of Geological and Planetary Sciences, California Institute of Technology, 1200 East California Blvd, Pasadena, CA 91125, USA}

\author[0000-0002-5375-4725]{Heather~A.~Knutson}
\affil{Division of Geological and Planetary Sciences, California Institute of Technology, 1200 East California Blvd, Pasadena, CA 91125, USA}

\author[0000-0002-2248-3838]{Leonardo~A.~dos~Santos}
\affil{Space Telescope Science Institute, 3700 San Martin Drive, Baltimore, MD 21218, USA}
\affil{Observatoire astronomique de l’Université de Genève, Chemin Pegasi 51, 1290 Versoix, Switzerland}

\author[0000-0002-6540-7042]{Lile~Wang}
\affil{Center for Computational Astrophysics, Flatiron Institute, New York, NY 10010, USA}

\author[0000-0002-8958-0683]{Fei~Dai}
\affil{Division of Geological and Planetary Sciences, California Institute of Technology, 1200 East California Blvd, Pasadena, CA 91125, USA}

\begin{abstract}
Observations of present-day mass-loss rates for close-in transiting exoplanets provide a crucial check on models of planetary evolution. One common approach is to model the planetary absorption signal during the transit in lines like He I 10830 with an isothermal Parker wind, but this leads to a degeneracy between the assumed outflow temperature $T_0$ and the mass-loss rate $\dot{M}$ that can span orders of magnitude in $\dot{M}$. In this study, we re-examine the isothermal Parker wind model using an energy-limited framework. We show that in cases where photoionization is the only heat source, there is a physical upper limit to the efficiency parameter $\varepsilon$ corresponding to the maximal amount of heating. This allows us to rule out a subset of winds with high temperatures and large mass-loss rates as they do not generate enough heat to remain self-consistent. To demonstrate the utility of this framework, we consider spectrally unresolved metastable helium observations of HAT-P-11b, WASP-69b, and HAT-P-18b. For the former two planets, we find that only relatively weak ($\dot{M}\lesssim 10^{11.5}$ g s$^{-1}$) outflows can match the metastable helium observations while remaining energetically self-consistent, while for HAT-P-18b all of the Parker wind models matching the helium data are self-consistent. Our results are in good agreement with more detailed self-consistent simulations and constraints from high-resolution transit spectra.
\end{abstract}


\section{Introduction} \label{sec:intro}
A majority of the extrasolar planets discovered by transit surveys orbit close to their host stars and are subjected to intense irradiation. The incident flux received by these planets can remove their atmospheres if the heating is large compared to the planet's gravitational potential. Indeed, we see evidence for atmospheric escape in the radius-period distribution of close-in planets (the `evaporation valley') observed by \textit{Kepler} and \textit{K2} \citep{Fulton17, Fulton18, vanEylen18, HardegreeUllman20}. We can quantify the present-day mass loss rates of the most favorable transiting planets by measuring the amount of absorption during the transit in spectral lines where outflowing material becomes opaque. To date, atmospheric outflows have been detected using Lyman-$\alpha$ \citep[e.g.][]{vidalMadjar03}, H$\alpha$ \citep[e.g.][]{Yan18}, metal lines \citep[e.g.][]{vidalMadjar04}, and metastable helium \citep[e.g.][]{Spake18}. 

The upper atmospheres of planets are engines that turn light into heat. This heating drives hydrodynamic escape by lifting material out of the planet's gravitational potential well. We can use this concept to calculate an ``energy-limited'' mass loss rate, as first done (albeit in the context of highly conductive atmospheres) by \citet{Watson81}: 

\begin{equation}
    \dot{M} = \frac{\varepsilon \pi R_\mathrm{XUV}^2 F_\mathrm{XUV}}{KGM_\mathrm{p}/R_\mathrm{p}}.
    \label{elim}
\end{equation}

The numerator of this equation is an estimate for the heating rate of the planet (in erg s$^{-1}$), where the stellar high-energy flux $F_\mathrm{XUV}$ (in erg s$^{-1}$ cm$^{-2}$) impinges on a cross-sectional area $\pi R_\mathrm{XUV}^2$ of the planet, heating it with efficiency $\varepsilon$. The denominator is the gravitational potential (in erg g$^{-1}$), with a correction factor $K$ to account for the fact that atoms need only be lifted past the Roche lobe to escape the planet \citep{Erkaev07}:

\begin{equation}
    K = 1 - \frac{3}{2}\Big(\frac{R_\mathrm{p}}{R_\mathrm{Roche}}\Big) + \frac{1}{2}\Big(\frac{R_\mathrm{p}}{R_\mathrm{Roche}}\Big)^3.
    \label{erkaevK}
\end{equation}
For small planet-to-star mass ratios, the Roche radius can be written as:

\begin{equation}
    R_\mathrm{roche} \approx a\Big(\frac{M_\mathrm{p}}{3M_\star}\Big)^{1/3}.
\end{equation}

The limitations of this formalism are well-documented \citep[e.g.][]{MurrayClay09, Salz16a, Salz16b, Kubyshkina18, Owen19, Krenn21}. Detailed hydrodynamical models show that the efficiency $\varepsilon$ depends strongly on a number of factors, including the planetary mass, radius, atmospheric composition, assumed heating processes, and stellar spectrum \citep{Owen12, Shematovich14}. Despite these drawbacks, this equation offers a convenient way to predict mass loss without expensive computational modeling, and it has been used extensively to model the \textit{Kepler} evaporation valley \citep{Lopez13, Owen13, Owen17}. 

The energy-limited framework has also recently been used to interpret measurements of metastable helium absorption from close-in transiting gas giant planets \citep{Lampon21a, Lampon21b}. Helium observations are typically modeled using an isothermal Parker wind (parameterized by an assumed outflow temperature $T_0$ and mass-loss rate $\dot{M}$) coupled to a set of photoionization equations that can be solved for the He level populations as a function of altitude \citep{Oklopcic18}. The photoionized Parker wind model is faster to compute than expensive self-consistent 3D simulations \citep{Wang21a, Wang21b}, making it easier to map out the parameter spaces that fit measurements (and non-detections) of absorption in the metastable helium triplet \citep{Mansfield18, Gaidos20a, Gaidos20b, Hirano20, Ninan20, Vissapragada20, Krishnamurthy21, Paragas21}. However, there is a degeneracy between $\dot{M}$ and $T_0$ in the Parker wind model, which can lead to uncertainties in $\dot{M}$ that span multiple orders of magnitude. This is particularly problematic for lower signal-to-noise and/or spectroscopically unresolved observations of helium absorption where the temperature of the outflow is unconstrained by the data. 

In this work, we show that the $\dot{M}-T_0$ degeneracy in helium observations can be partially resolved because there is an upper limit to the efficiency of an energy-limited H/He outflow corresponding to the maximal amount of heating from photoionization. In \S\ref{sec:elim}, we derive an expression for $\varepsilon_\mathrm{max}$ in a lossless isothermal wind, i.e. a wind in which radiative cooling is negligible. We show that this expression can be used to define a bounded region in the $\dot{M}-T_0$ plane, beyond which there is not enough heat to power the outflow. In \S\ref{sec:w69} we use this framework to resolve the $\dot{M}-T_0$ degeneracy in previous modeling of metastable helium absorption from HAT-P-11b, WASP-69b, and HAT-P-18b. Finally, we offer some concluding thoughts in \S\ref{sec:conc}.

\section{The Maximum Mass-Loss Efficiency} \label{sec:elim}
We first calculate the maximal outflow efficiency $\varepsilon$ for a lossless isothermal wind, assuming that the heat source is photoionization of H and He. Then, we use the maximum efficiency to trace out a critical region on the $\dot{M}-T_0$ plane, which can be used to assess energetic self-consistency. Because constants can always be absorbed into the efficiency $\varepsilon$, we will make the common arbitrary definition $R_\mathrm{XUV} \equiv R_\mathrm{p}$ in Equation~(\ref{elim}) and absorb the constant $(R_\mathrm{XUV}/R_\mathrm{p})^2$ into $\varepsilon$, admitting the somewhat awkward possibility that $\varepsilon > 1$ if $R_\mathrm{XUV} \gg R_\mathrm{p}$. We note that other authors have proceeded by fixing $\varepsilon$ and calculating $R_\mathrm{XUV}$ instead \citep[e.g.][]{Erkaev07, Salz16b, Kubyshkina18}. This calculation could be restructured for $R_\mathrm{XUV}$ or the product $\varepsilon R_\mathrm{XUV}^2$, but we choose to work with the efficiency term because this is typically the free parameter assumed in planetary population studies. We also note that varying authors have different definitions of the ``efficiency'' term in the context of planetary wind. For example, \citet{Shematovich14} defines it as the ratio of the local heating rate to the radiative input (treating the cooling processes separately), which is adopted by \citet{Erkaev07} and \citet{Kubyshkina18}, whereas \citet{Salz16a} includes the cooling rate explicitly. Throughout this work, when we refer to the ``efficiency,'' we are referring explicitly to the $\varepsilon$ term in Equation~(\ref{elim}).

\subsection{An Upper Limit to the Mass-Loss Efficiency}
To compute $\varepsilon$, we begin by writing down the equation for energy balance in the wind \citep{Lamers99, Erkaev07}:

\begin{equation}
    \dot{M}\Big(\frac{KGM_\mathrm{p}}{R_\mathrm{p}} + \frac{\Delta v^2}{2} + \frac{5}{2}\frac{k\Delta T}{\mu m_\mathrm{H}}\Big) = \iiint \limits \Gamma_\mathrm{net} dV,
\end{equation}
where $\dot{M}$ is the mass-loss rate of the planet, $K$ is as defined in Equation~(\ref{erkaevK}), $\Delta v$ is the difference in outflow velocity between $R_\mathrm{p}$ and $R_\mathrm{Roche}$, $\Delta T$ is the difference in temperature between $R_\mathrm{p}$ and $R_\mathrm{Roche}$, and $\mu$ is the mean particle mass of the outflow. The terms on the left-hand side correspond to the difference in gravitational potential energy, kinetic energy, and enthalpy (for a monatomic outflow) between the planetary radius and the Roche lobe. On the right-hand side is the integrated net heating rate per unit volume, with $\Gamma_\mathrm{net} = \Gamma - \Lambda$ (where $\Gamma$ and $\Lambda$ are the heating and cooling rates, respectively) in erg s$^{-1}$ cm$^{-3}$. 

In this work we only consider isothermal winds where $\Delta T$ = 0 and assume that the kinetic energy term is small. In the Parker wind model, material is accelerated to velocities on the order of the sound speed at the Roche lobe. This means that $\Delta v \sim 10^{6}$~cm s$^{-1}$ \citep[e.g.][]{MurrayClay09, Oklopcic18} and $(\Delta v)^2/2 \sim 5\times10^{11}$~erg/g, whereas the gravitational potential term is $\gtrsim 10^{12}$~erg/g for all of the planets considered in this work. These are the same assumptions made by \citet{Erkaev07} to reproduce the energy-limited mass loss expression from \citet{Watson81} in Equation~(\ref{elim}).

With these assumptions, and using Equation~(\ref{elim}) to substitute for the gravitational potential term, we have:

\begin{equation}
    \varepsilon \pi R_\mathrm{p}^2F_\mathrm{XUV} = \iiint \limits (\Gamma - \Lambda) dV,
\end{equation}
This simply reflects our assumption that the numerator of Equation~(\ref{elim}) is a proxy for the total heating experienced by the planet. We then calculate $\varepsilon$ for an atmosphere where radiative losses are negligible; that is, one where all heating goes into driving the outflow and $\Lambda \ll \Gamma$. This corresponds to: 

\begin{equation}
    \varepsilon_\mathrm{max} \pi R_\mathrm{p}^2F_\mathrm{XUV} = \iiint \limits_H \Gamma dV, \label{proxy}
\end{equation}
where $\varepsilon_\mathrm{max}$ is the maximum mass-loss efficiency and $H$ denotes the dayside hemisphere where the heating occurs. If we assume that all heating comes from photoionization, we can write the total heating rate as a sum of heating rates for each photoprocess of interest \citep[e.g.][]{Osterbrock06}:

\begin{align}
    \Gamma &= \sum_i\Gamma_i \\
    &= \sum_i n_{i,r}\int_{\min(\nu_i)}^\infty (h\nu - h\nu_i) \nonumber \\
    &\times \frac{F_\nu \cos(\theta) \exp(-\tau_{\nu, r})}{h\nu}\sigma_{\nu,i} d\nu.
\end{align}

In this equation, $h\nu$ denotes the photon energy. The subscript $i$ denotes different photoproceses (for instance, hydrogen photoionization), the subscript $\nu$ denotes a quantity that varies with frequency, and the subscript $r$ denotes one that varies with radius. The number density profile of the species being photoionized is $n_{i,r}$, the ionization threshold for the process is $h\nu_i$, and $h\nu - h\nu_i$ is the maximum yield per photoprocess. For example, a photon with energy $h\nu = 20$ eV yields 6.4~eV when ionizing a hydrogen atom out of its ground state ($h\nu_\mathrm{H}$ = 13.6 eV). $F_\nu \cos(\theta) \exp(-\tau_{\nu, r})$ is the spectrum of radiation reaching radius $r$ from the center of the planet and latitude $\theta$ with respect to the substellar point in erg s$^{-1}$ cm$^{-2}$ Hz$^{-1}$. Finally, $\sigma_{\nu,i}$ is the frequency-dependent cross-section for photoprocess $i$. The frequency integral is taken from the minimum photoionization threshold over all relevant photoprocesses $\min(\nu_i)$ to allow for a consistent definition of $F_\mathrm{XUV}$. 

The optical depth term is defined as:

\begin{equation}
    \tau_{\nu,r} = \sum_i \sigma_{i,\nu}\int_r^\infty n_{i,r} dr.
\end{equation} 

This definition implictly assumes that the optical depth reaches a large value at $R_\mathrm{p}$. For high-energy photons where the photoionization cross-sections are small this condition may not always be satisfied, implying that photons can freely reach this depth. However, these photons cannot pass through the planet, and their energy must be deposited somewhere. We assume that they participate in the energy balance of the lower atmosphere at $R \sim R_\mathrm{p}$, and that they do not heat the outflow.

We collect the photon energy terms into a normalized yield term $\eta_\nu$:

\begin{equation}
    \eta_\mathrm{\nu} = \frac{h\nu - h\nu_i}{h\nu} = 1 - \frac{\nu_i}{\nu}.
    \label{yield}
\end{equation}
In reality, this is an upper limit on the yield. We have assumed that the photoelectrons resulting from ionization transform all of their energy into heat, but they can also excite or ionize other atoms \citep{MurrayClay09, Shematovich14}. By neglecting these minor energy deposition channels, we obtain a slightly more generous upper limit than we would otherwise. With the yield term now defined, we can write the total heating rate as:

\begin{equation}
    \Gamma = \sum_i n_{i,r}\int_{\min(\nu_i)}^\infty \eta_\nu F_\nu\cos(\theta)\exp(-\tau_{\nu, r})\sigma_{\nu,i} d\nu. 
    \label{heatrate}
\end{equation}

We integrate this equation over the dayside hemisphere \citep[for the geometry of the integral see e.g. Chapter 2 of][]{Seager10}:

\begin{align}
    \iiint \limits_H \Gamma dV &= \int_0^{2\pi}\int_0^{\pi/2}\int_{R_\mathrm{p}}^{R_\mathrm{Roche}} \Gamma r^2\sin\theta dr d\theta d\phi \label{integral} \\
    &= \pi\int_{R_\mathrm{p}}^{R_\mathrm{Roche}} r^2 \sum_i n_{i,r}\nonumber \\
    &\times \int_{\min(\nu_i)}^\infty \eta_\nu F_\nu\exp(-\tau_{\nu, r})\sigma_{\nu,i} d\nu dr. \label{integrated}
\end{align}
We then substitute Equation~(\ref{integrated}) into Equation~(\ref{proxy}), where the factors of $\pi$ cancel. Noting that $F_\mathrm{XUV}$ can be written as:

\begin{equation}
    F_\mathrm{XUV} = \int_{\min(\nu_i)}^\infty F_\nu d\nu,
\end{equation}
we arrive at:

\begin{align}
    \varepsilon_\mathrm{max} &=  \frac{1}{R_\mathrm{p}^2} \int_{R_\mathrm{p}}^{R_\mathrm{Roche}} r^2 \nonumber \\
    &\times \sum_i n_{i,r}\frac{\int_{\min(\nu_i)}^\infty \eta_\nu F_\nu\exp(-\tau_{\nu, r})\sigma_{\nu,i} d\nu}{\int_{\min(\nu_i)}^\infty F_\nu d\nu} dr. \label{fullexpr}
\end{align}

\subsection{Heating Cross-Sections}
We can define the ratio of frequency integrals in Equation~(\ref{fullexpr}) as a cross section weighted by the stellar spectrum and heating efficiency for photoprocess $i$. We call this the ``heating cross-section'', and note that the optical depth dependence causes it to vary with radius:

\begin{equation}
    \bar{\sigma}_{i,r} = \frac{\int_{\min(\nu_i)}^\infty \eta_\nu F_\nu\exp{(-\tau_{\nu,r})}\sigma_{\nu,i} d\nu}{\int_{\min(\nu_i)}^\infty F_\nu d\nu} \label{specavg}
\end{equation}

To calculate this heating cross-section, we require a stellar spectrum and a frequency-dependent cross-section for each photoprocess. For the spectrum, we use the v2.2 panchromatic SEDs at 1~\AA~binning from the MUSCLES survey \citep{France16, Loyd16, Youngblood16}. For the hydrogen photoionization cross-section, we use \citep[e.g.][]{Osterbrock06}:

\begin{align}
    \sigma_i(\nu) &= \sigma_{\nu_i}\frac{\exp\Big(4 - 4\frac{\tan^{-1}\epsilon}{\epsilon}\Big)}{1 - \exp{(-2\pi / \epsilon)}}\Big(\frac{\nu_i}{\nu}\Big)^4, \nonumber \\
    &\qquad \nu > \nu_i,
    \label{hcross}
\end{align}
where $\epsilon = \nu/\nu_\mathrm{i} - 1$. The threshold energy for neutral hydrogen photoionization is $h\nu_\mathrm{H} = 13.6$~eV, and the cross-section at this threshold is $\sigma_{\nu_\mathrm{H}} = 6.3\times10^{-18}$~cm$^{-2}$. We also consider heat generated by the photoionization of ionized helium, a relatively minor process. The cross section for this hydrogen-like species is also given by Equation~(\ref{hcross}), but the threshold energy is $h\nu_\mathrm{He^{+}} = Z^2h\nu_\mathrm{H} = 54.4$~eV, and the cross-section at this threshold is $\sigma_{\nu_\mathrm{He^{+}}} = \sigma_{\nu_\mathrm{H}}/Z^2 = 1.6\times10^{-18}$~cm$^{-2}$. 
For the neutral helium photoionization cross-section, we use \citep{Yan98}:

\begin{align}
    \sigma_{\mathrm{He}}(\nu) &= \frac{\sigma_{\mathrm{a}}}{(h\nu / 1~\mathrm{keV)^{7/2}}}\Bigg(1 + \sum_{n = 1}^{6}\frac{a_n}{(\nu/\nu_\mathrm{He})^{n/2}}\Bigg), \nonumber \\
    &\qquad \nu > \nu_\mathrm{He}
    \label{hecross}
\end{align}
where the threshold photon energy $h\nu_\mathrm{He} = 24.6$~eV, $\sigma_{\mathrm{a}} = 7.33\times10^{-22}$cm$^{2}$, and the constants $a_n$ are from Table~4 of \citet{Yan98}. 

With the heating cross-section defined, we can now write the expression for $\varepsilon_\mathrm{max}$ in a more illustrative form:

\begin{align}
    \varepsilon_\mathrm{max} =  \frac{1}{R_\mathrm{p}^2} \int_{R_p}^{R_\mathrm{Roche}} r^2 \sum_i n_{i,r}\bar{\sigma}_{i,r} dr. \label{finaleq}
\end{align}
The $n\sigma$ terms in the summation, which we will refer to as ``heating coefficients'', can be thought of analogously to absorption coefficients \citep[e.g.][]{Seager10}. Thus, the efficiency of an isothermal wind cannot be greater than the normalized second moment of its total heating coefficient.

We note that Equation~(\ref{finaleq}) gives us a simple way to compare the relative heating efficiencies of neutral hydrogen and helium. These processes will be similarly important for heating the atmosphere when the heating coefficients are comparable: $n_\mathrm{H}\bar\sigma_\mathrm{H} \sim n_\mathrm{He}\bar\sigma_\mathrm{He}$. In the high-energy limit, the cross-section for neutral helium photoionization is an order of magnitude larger than that for hydrogen photoionization, but there are also fewer helium atoms to ionize. We consider a 90-10 hydrogen-helium atmosphere that is optically thin, i.e. where $\exp(-\tau_{\nu,r}) \sim 1$ in Equation~(\ref{specavg}). In this case, the heating coefficients depend only on the assumed stellar spectrum. Stepping through the ensemble of stellar spectra collected by MUSCLES, we find that the optically-thin heating coefficient for helium ranges from 20\% (for $\epsilon$ Eri) to 67\% (for GJ 1214) that of hydrogen. For a stellar spectrum similar to the M stars in the MUSCLES sample, helium photoionization can be a large (though still sub-dominant) heat source; pure-hydrogen escape models that have been developed for planets orbiting earlier stars may therefore underestimate the total thermospheric energy budget for planets orbiting M stars. K stars in this sample output relatively more flux near the hydrogen photoionzation threshold, so helium photoionization is a smaller heat source for planets orbiting K stars. These relative efficiencies are similar to those obtained from more detailed computational modeling \citep{Salz16b}. These calculations are only meant to be illustrative; in general, the optical depth term $\exp(-\tau_{\nu,r})$ can be important, and the heating coefficients change substantially with radius.

\subsection{Energetic Self-Consistency}
We next consider the implications of this upper limit on the mass loss efficiency for a photoionized isothermal Parker wind model, which parameterizes the density structure at the substellar point $n_{i,r}$ in terms of a mass-loss rate $\dot{M}$ and the outflow temperature $T_0$. For helium studies, the Parker wind is typically taken to be one-dimensional, but this greatly overestimates the work done. The wind is strongest at the substellar point because the irradiation is strongest there, but decreases in strength towards the terminator due to the diminished incident flux, and no wind is launched on the planetary nightside as the atmosphere is not irradiated there. This can be accounted for as follows: the planet is heated over only $\pi$ steradians (Equation~\ref{integrated}), so the outflow is driven only over $\pi$ steradians on the planetary dayside and the work done is decreased by a factor of 1/4 compared to a 1D isotropically-irradiated outflow \citep{Stone09, Salz16a}.

Because the density distributions are functions of $\dot{M}$ and $T_0$ in this model, the energy-limited efficiency is itself a function of these parameters, and the mass-loss rate must satisfy:
\begin{equation}
    \frac{\dot{M}}{4} \leq \frac{\varepsilon_\mathrm{max}(\dot{M}, T_0)\pi R_\mathrm{p}^3 F_\mathrm{XUV}}{KGM_\mathrm{p}},
    \label{consistency}
\end{equation}
where the factor of 1/4 comes from the aforementioned 3D outflow geometry. Had we derived $\varepsilon_\mathrm{max}$ assuming an isotropically-irradiated 1D outflow without this geometric factor, then we would be integrating over the entire planet in Equation~(\ref{integral}), picking up a factor of 4 (the angular part of the integral would evaluate to 4$\pi$ rather than $\pi$). Thus, the constraint on $\dot{M}$ would remain the same.

This inequality defines an allowed region on the $\dot{M}-T_0$ plane. The problem is fundamentally one of energetic self-consistency: if we assume a mass-loss rate and thermosphere temperature, do the resulting density profiles allow enough heating to power the assumed mass-loss rate? For solutions outside the allowed region, the answer is no: they are not permitted unless there is some additional source of heat. Solutions on the boundary are exactly energy-limited; that is, they generate exactly enough heat to power their outflow. Solutions inside the boundary generate excess heat, which can always be balanced by additional cooling. If the cooling rate is known precisely, Equation~(\ref{consistency}) traces out a curve rather than an allowed region. We briefly consider this case in Appendix~\ref{cooling}.

\subsection{Regime of Validity}

We note that our calculation is valid only for gas giant planets with H/He-rich envelopes and lower gravitational potentials experiencing (approximately) radially symmetric outflows. Detailed numerical simulations indicate that planets with gravitational potentials $\Phi_\mathrm{p} \gtrsim 10^{13.1}$~erg~g$^{-1}$ exhibit strong temperature gradients, and cooling through Lyman-$\alpha$ and free-free emission becomes important \citep{Salz16a, Salz16b}. We have focused on isothermal, lossless outflows, so our model is therefore better suited to modeling lower gravity planets. High-gravity planets will still be governed by energy balance, but the isothermal Parker wind will not correctly describe the density distributions. We also note that H/He-rich Parker winds have been inadequate in predicting metastable helium signatures for sub-Neptunes thus far, which may indicate these planets differ substantially in composition from the pure H/He atmosphere assumed here \citep{Kasper20, Krishnamurthy21}. Therefore, we do not recommend using this methodology for planets with $R_\mathrm{p} \lesssim 4R_\Earth$. These criteria are satisfied for $\sim30\%$ of the $\sim$1000 transiting exoplanets with measured masses and radii on the NASA Exoplanet Archive.

Our model additionally assumes an idealized geometry for the wind, and will be less accurate for outflows deviating from the assumed geometry. There are several factors that contribute to such deviations. Interactions between the planetary and stellar winds can sculpt the outflow into a comet-like tail \citep{McCann19, Wang21b, MacLeod21}, and we therefore expect the model to perform worse for planets orbiting young and/or exceptionally active stars. WASP-107 b orbits a relatively active star and has a strongly blueshifted helium line profile and strong post-egress absorption \citep{Allart19, Kirk20, Spake21}, both signs of an asymmetric outflow. There were initially hints of a similar tail for WASP-69b \citep{Nortmann18}. However, subsequent observations and modeling both indicate that this planet's outflow is relatively symmetric \citep{Vissapragada20, Wang21a}. Simulations and observations of the transiting planet HD 63433c, which orbits a 440 Myr old star, indicate that it also has a comet-like tail \citep{Zhang21}, and simulations of the young planet AU Mic b suggest the strong stellar wind should shape its outflow as well \citep{Carolan20}, though helium has not yet been conclusively detected in its atmosphere \citep{Hirano20}. 

Magnetically-controlled outflows are also expected to deviate from the idealized geometry in this work. Outflowing material near equatorial latitudes follow closed magnetic field lines falling back onto the planet, whereas material near the poles may escape \citep{Adams11, Trammell11, Owen14, Trammell14}. However, it is presently unclear to what extent magnetic fields may affect metastable helium observations, which have thus far been well-fit by models without magnetic fields. This may be because the planetary outflows are dominated by ram pressure rather than magnetic pressure within the metastable helium photosphere \citep{Zhang21}.

\section{Application to Metastable Helium Observations}
\label{sec:w69}
To show how the constraints from our energy-limited model may be used in practice, we consider published observations of metastable helium absorption from HAT-P-11b, WASP-69b, and HAT-P-18b. These three planets have all been observed using narrowband photometry or low-resolution spectroscopy, which provide minimal information about the line shape. This means that when an isothermal Parker wind model is used to fit the measured absorption signal, there is a large degeneracy between the mass-loss rate and the assumed outflow temperature. This degeneracy can be broken with line-shape measurements of sufficiently high precision, which independently constrain the outflow temperature \citep[e.g.][]{dosSantos21}. However, the line shape can also vary when the absorbing region is not optically thin \citep{Salz18}, and when the outflow itself kinematically broadens the line \citep{Allart19, Wang21b, Seidel21}. This means that line shape alone is an imperfect proxy for outflow temperature. Even if this was not the case, the signal-to-noise ratio of the measured line profile is often too low to provide a useful constraint on the outflow temperature \citep{Lampon20, Lampon21a}. 

In this section, we show that we can use the energy-limited framework to place an upper bound on the temperatures of these outflows, resulting in tighter constraints on the retrieved mass-loss rates from the Parker wind model. For each of the three planets considered here, we used the \texttt{p-winds} code \citep[e.g.][]{dosSantos21} to calculate isothermal Parker wind models and compared those models to the observed metastable helium signal. \texttt{p-winds} is an open-source implementation of the model described by \citet{Oklopcic18} and \citet{Lampon20}. We evaluated these models over a grid defined by $\dot{M} = 10^9 - 10^{12}$~g~s$^{-1}$ and $T_0 = 5000 - 15000$~K. The temperature grid roughly corresponds to the range of temperatures seen in self-consistent simulations \citep{Salz16a, Wang21a, Wang21b}. At each point on our grid, we computed the outflow density structure (for a 90/10 H/He composition) and calculated the corresponding ionization structure and level populations with \texttt{p-winds}. In these calculations, we noticed that the approximation for the hydrogen photoionization rate in Equation~(9) of \citet{Oklopcic18} tended to over-predict the rate close to the ionization front, which is consequential for the modeled mass-loss efficiency. We therefore updated the calculation in \texttt{p-winds} to avoid the approximation. We also note that our number density of ionized helium is overestimated because some of the ionized helium will end up populating He$^{2+}$ in steady-state, which \texttt{p-winds} does not account for. By ignoring this minor \citep[e.g.][]{Salz16a} sink, we obtain a slightly larger $\varepsilon_\mathrm{max}$ than we otherwise would.

We then used the framework from \S\ref{sec:elim} to calculate the self-consistent $\dot{M}-T_0$ region for each planet. At each $(\dot{M}, T_0)$ wind on the grid, we integrated the density distributions of H and He to obtain column densities, from which we computed the total optical depth as a function of frequency and radius $\tau_{\nu,r}$. Using this optical depth, the cross sections in Equations~(\ref{hcross}) and (\ref{hecross}), the yield term in Equation~(\ref{yield}), and the assumed stellar spectrum, we then calculated the heating cross-sections using Equation~(\ref{specavg}). Finally, we multiplied by the density distributions to get the heating coefficients, summed to get the total heating coefficient, and took the second moment to get the upper limit on the mass-loss efficiency per Equation~(\ref{finaleq}). To assess whether or not a given outflow was self-consistent, we used this $\varepsilon_\mathrm{max}$ to calculate the maximum mass-loss rate via Equation~(\ref{consistency}). If we found a maximum mass-loss rate lower than the assumed mass-loss rate, the solution was energetically inconsistent; otherwise, it was admissible. This procedure has been added in v1.2.4 of the \texttt{p-winds} code.

\subsection{HAT-P-11b} \label{sec:hp11}

\begin{figure}[ht!]
\centering
\includegraphics[width=0.5\textwidth]{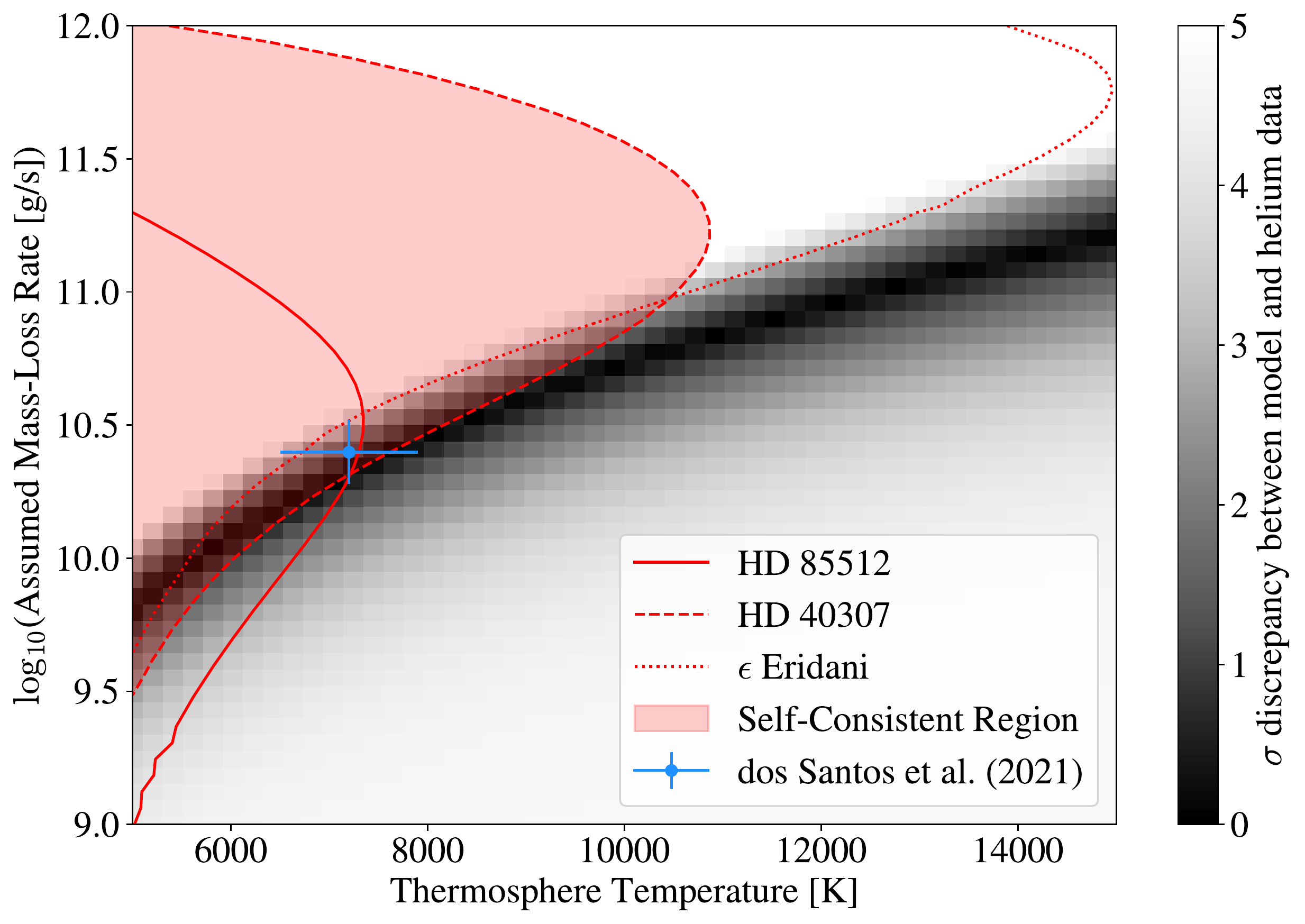}
\caption{Mass-loss modeling for the HAT-P-11b observations presented in \citet{Mansfield18}. The black shading indicates the $\sigma$ discrepancy between the mass-loss model in each grid cell and the metastable helium observation. The red shading indicates the self-consistent region from Equation~(\ref{consistency}) using the scaled spectrum of HD~40307, and the boundary for a perfectly energy-limited outflow assuming this spectrum is given with the dashed red line. The solid and dotted red lines give the boundary of the self-consistent regions for HD 85512 and $\epsilon$ Eridani, respectively. The blue point additionally indicates the 1$\sigma$ range from the \citet{dosSantos21} retrieval on the \citet{Allart18} high-resolution spectrum.}
\label{hp11}
\end{figure}

HAT-P-11b is slightly larger in mass and radius than Neptune ($\Phi_\mathrm{p} \approx 3 \times 10^{12}$~erg~g$^{-1}$), and orbits its K4 host star with a period of 5 days \citep{Bakos10}. \citet{Mansfield18} detected helium absorption in the atmosphere of this planet at low resolving power with \textit{HST} WFC3/G102, and \citet{Allart18} detected a comparable absorption signal at high resolving power with CARMENES. For this planet, the high signal-to-noise of the line shape measured by \citet{Allart18} provides us with an independent measurement of the outflow temperature. \citet{dosSantos21} fitted the line absorption profile from \citet{Allart18} with a Parker wind model using the \texttt{p-winds} code and found that it was best matched by an outflow with $T_0 = 7200\pm700$~K and $\dot{M} = 2.5^{+0.8}_{-0.6}\times10^{10}$~g~s$^{-1}$. In this study we instead fitted the unresolved measurements from \citet{Mansfield18}, and used our energy-limited framework to help resolve the $\dot{M}-T_0$ degeneracy. We use the stellar spectrum of HD 40307 for this calculation, which is the same choice made by \citet{dosSantos21}.

We show the resulting constraints on the outflow rate and temperature in Figure~\ref{hp11}, and compare these constraints to the result from \citet{dosSantos21}. We find that some of the previously reported Parker wind solutions with high temperatures and high mass-loss rates are in fact energetically inconsistent; i.e. they do not have enough H and He to power the outflow by photoionization. This allows us to partially resolve the degeneracy between mass-loss rate and thermosphere temperature: the outflow must be cooler than $\sim 10,000$~K and relatively weak ($\dot{M} \lesssim 10^{11}$~g~s$^{-1}$). Our results are in good agreement with \citet{dosSantos21}, and their retrieved value matches our energy-limited contour within the 1$\sigma$ level. We therefore conclude that this outflow is energy-limited. This is also consistent with the simulations of \citet{Salz16a}, who explicitly modeled radiative cooling in HAT-P-11b's outflowing atmosphere and found it to be negligible. 

\begin{figure}[ht!]
\centering
\includegraphics[width=0.5\textwidth]{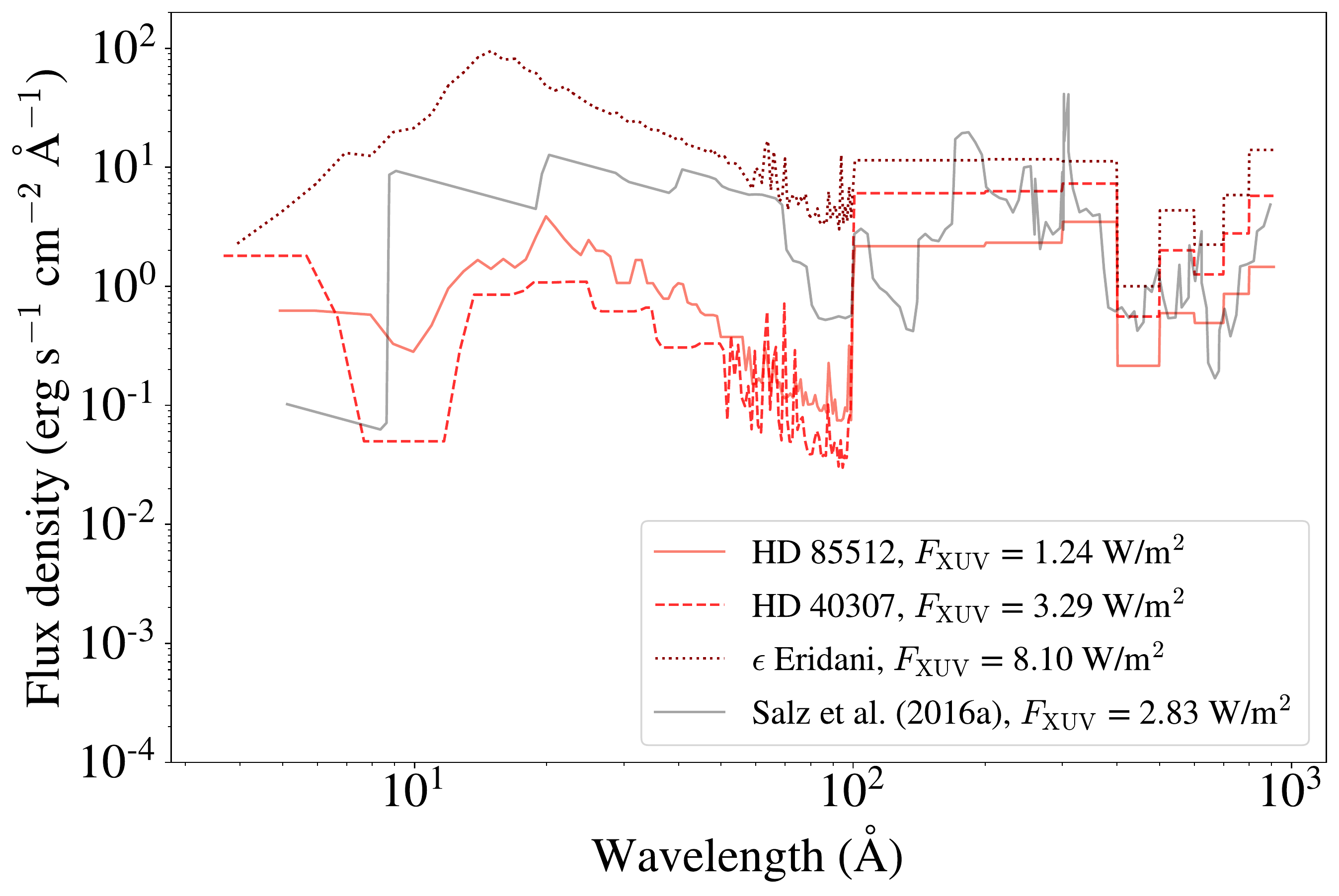}
\caption{High-energy stellar spectra considered for HAT-P-11. All spectra have been scaled to the location of HAT-P-11b. Note that the \citet{Salz16a} spectrum is binned differently than the three MUSCLES spectra.}
\label{fig:spectra}
\end{figure}

We also quantify our sensitivity to the choice of high-energy stellar spectrum by overplotting the boundary of the self-consistent region calculated using scaled spectra for HD 85512, which is less energetic in the XUV than the HD 40307 spectrum, and $\epsilon$ Eridani which is more energetic in the XUV (see Figure~\ref{fig:spectra}). To be fully consistent, we also re-calculated the constraints from the Parker wind models with the same stellar spectrum. However, we found that this does not substantially change the range of mass-loss rates and temperatures that are consistent with the metastable helium data, as the stellar spectra have similar shapes at wavelengths longward of 100~\AA. In Figure~\ref{hp11}, we show that the range of self-consistent solutions is somewhat larger when we use $\epsilon$ Eridani as a proxy for the star's high energy spectrum, and smaller when we use HD 85512, reflecting the differences in XUV luminosities for these stars. A larger XUV intensity tends to increase ionization throughout the outflow \citep[though this is also somewhat dependent on the spectral shape, see e.g.][]{Guo16}, leading to lower neutral densities, and thus smaller values of $\varepsilon_\mathrm{max}$ per Equation~(\ref{finaleq}). However, this effect is sub-linear and does not compensate for the linear dependence on $F_\mathrm{XUV}$ in the expression for the maximum mass-loss rate in Equation~(\ref{consistency}), so larger XUV luminosities lead to larger self-consistent regions. 

\begin{figure}[ht!]
\centering
\includegraphics[width=0.5\textwidth]{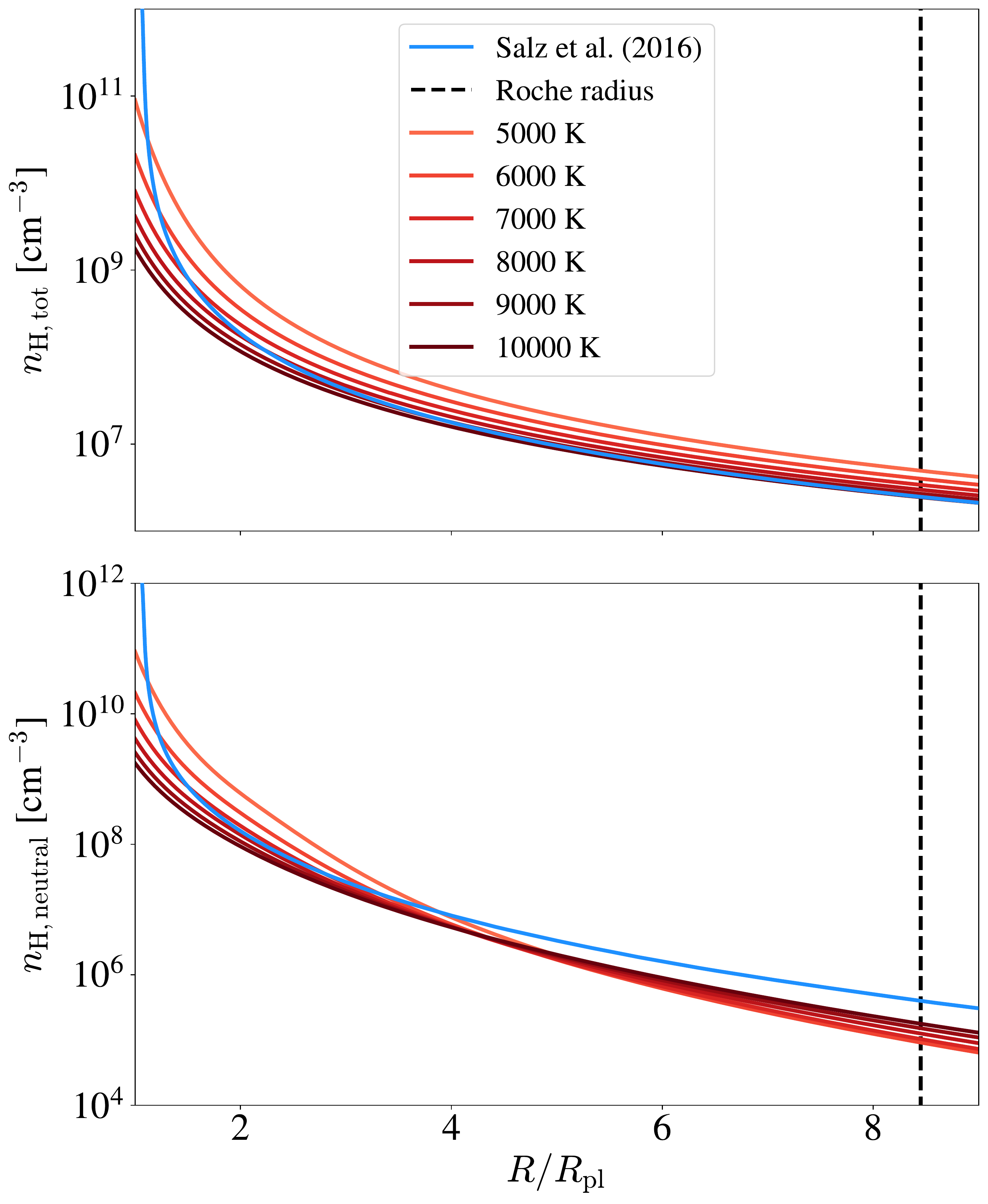}
\caption{Total number density of hydrogen (top) and number density of neutral hydrogen (bottom) for HAT-P-11b at the substellar point. Profiles from the \citet{Salz16a} simulation are shown in blue, and our photoionized Parker wind model in red, with darker colors corresponding to higher temperatures.}
\label{comparison_salz}
\end{figure}

We can use the \citet{Salz16a} simulation to explore the validity of the Parker wind model for this planet, as their model self-consistently predicts the temperature and density profiles at the substellar point. Their model suggests a mass-loss rate at the sub-stellar point of $10^{10.89}$~g~s$^{-1}$, which they divide by 4 to correct for the non-uniform radiation (as we do in Equation~\ref{consistency}) for a final mass-loss rate of $\dot{M}_\mathrm{sim} = 10^{10.29}$~g~s$^{-1}$. As we are comparing density profiles at the substellar point we selected isothermal Parker wind models with $\dot{M} = 10^{10.89}$~g~s$^{-1}$ and $T_0$ ranging from 5000~K to 10000~K in 1000~K steps. For the photoionization calculation we used the spectrum of HD 40307, which is closest to the spectrum used by \citet{Salz16a}. In Figure~\ref{comparison_salz}, we plot the total number density of hydrogen and the number density of neutral hydrogen from our models and compare it to the self-consistent simulations. 

Throughout most of the outflow, the total hydrogen number densities agree quite well between the two models. They differ by orders of magnitude in total density near $R = R_\mathrm{p}$, but this is expected. The Parker wind model is a poor approximation for the lower atmosphere, but this region has a negligible effect on the predicted outflow. The high number density of hydrogen close to the planet leads to a very large optical depth near $R_\mathrm{p}$ so it does not contribute much to the heating budget in either model. Our model appears to give somewhat smaller number densities for neutral hydrogen near the Roche radius. This likely arises from our use of a K star spectrum from MUSCLES, whereas \citet{Salz16a} used an inactive solar spectrum to reconstruct the shape of the spectrum in the EUV (see Figure~\ref{fig:spectra}). The discrepancy grows to about a factor of 2 at the Roche radius, but the densities here are small. Overall, the agreement between the two models is reasonable given the difference in methodology, and both models agree with the retrieval from \citet{dosSantos21} for HAT-P-11b.

\subsection{WASP-69b}
WASP-69b is a Neptune-mass, Jupiter-sized planet ($\Phi_\mathrm{p} \approx 4 \times 10^{12}$~erg~g$^{-1}$) in a 4~day orbit around a K5 host star \citep{Anderson14}. The low density and large scale height of this planet, combined with the host star's favorable ratio of EUV to XUV flux \citep{Oklopcic19}, made it an ideal target for initial studies of mass loss with the metastable helium line. The helium absorption signal for this planet has been measured both spectroscopically \citep{Nortmann18} and photometrically \citep{Vissapragada20}. In the latter work, we attempted to infer the mass-loss rate of WASP-69b using the Parker wind model from \citet{Oklopcic18}, and were able to place joint constraints on its mass-loss rate and thermosphere temperature. 

\begin{figure}[ht!]
\centering
\includegraphics[width=0.5\textwidth]{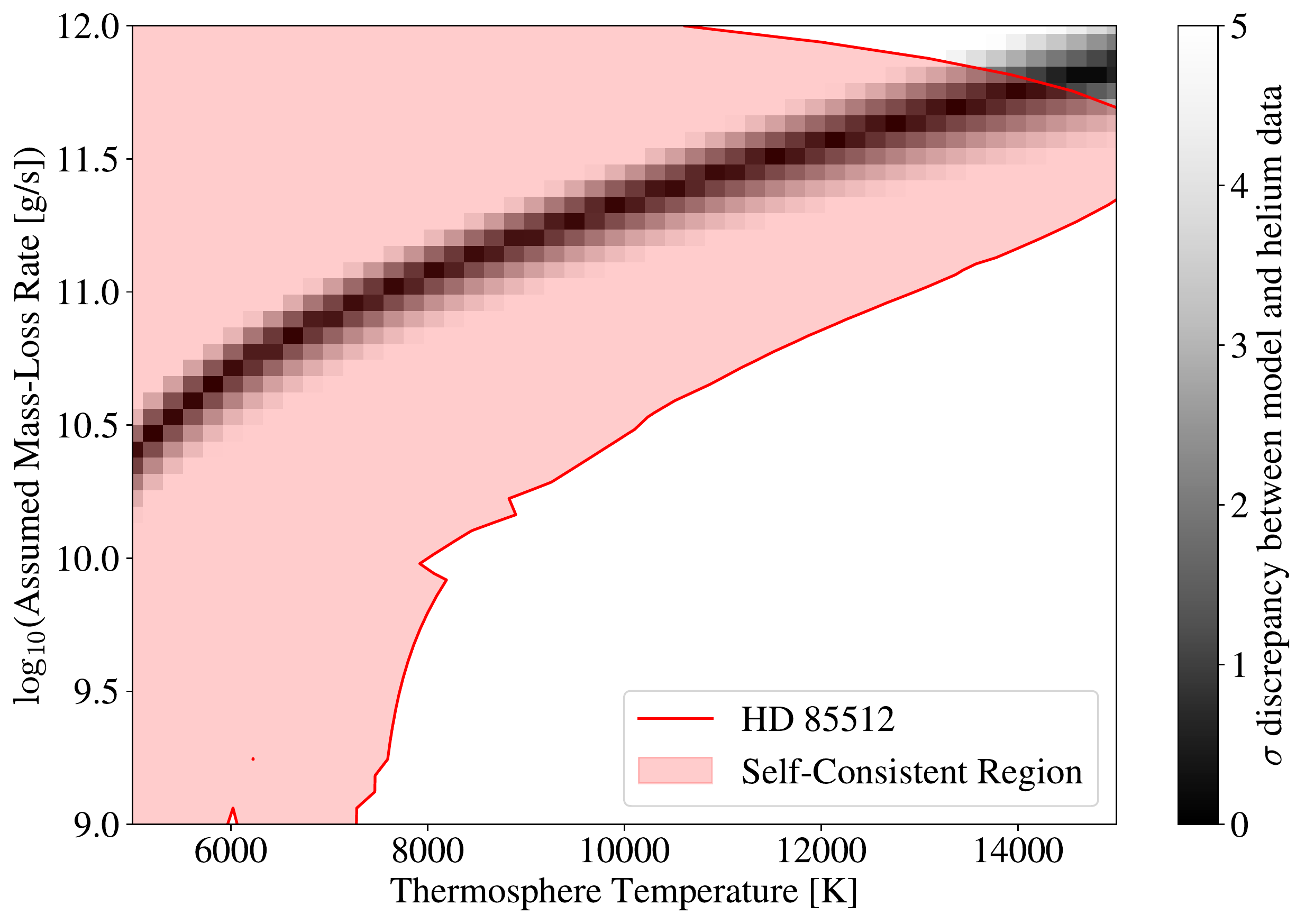}
\caption{Same as Figure~\ref{hp11}, but for the WASP-69b observations presented in \citet{Vissapragada20}.}
\label{wasp69}
\end{figure}

In Figure~\ref{wasp69}, we show the energetically self-consistency region and compare it to the set of Parker wind models that match the metastable helium absorption reported in \citet{Vissapragada20}, using the stellar spectrum of HD 85512 as we did in that work. We find that the outflow must be cooler than 14,000~K and relatively weak ($\dot{M} \lesssim 10^{11.5}$~g~s$^{-1}$). WASP-69b was also modeled self-consistently by \citet{Wang21a}, who used a more sophisticated 3D approach coupling hydrodynamics and thermochemistry. These authors found that the data were best-matched by models with $\dot{M}\sim10^{11}$~g~s$^{-1}$, in agreement with our upper limit. The \citet{Wang21a} model also achieves a maximum temperature in the substellar direction of $\sim 10^4$~K, so \citep[similarly to][]{Lampon20} we find that the isothermal Parker wind model agrees best with non-isothermal models near the maximum temperature.

\subsection{HAT-P-18b}
\begin{figure}[ht!]
\centering
\includegraphics[width=0.5\textwidth]{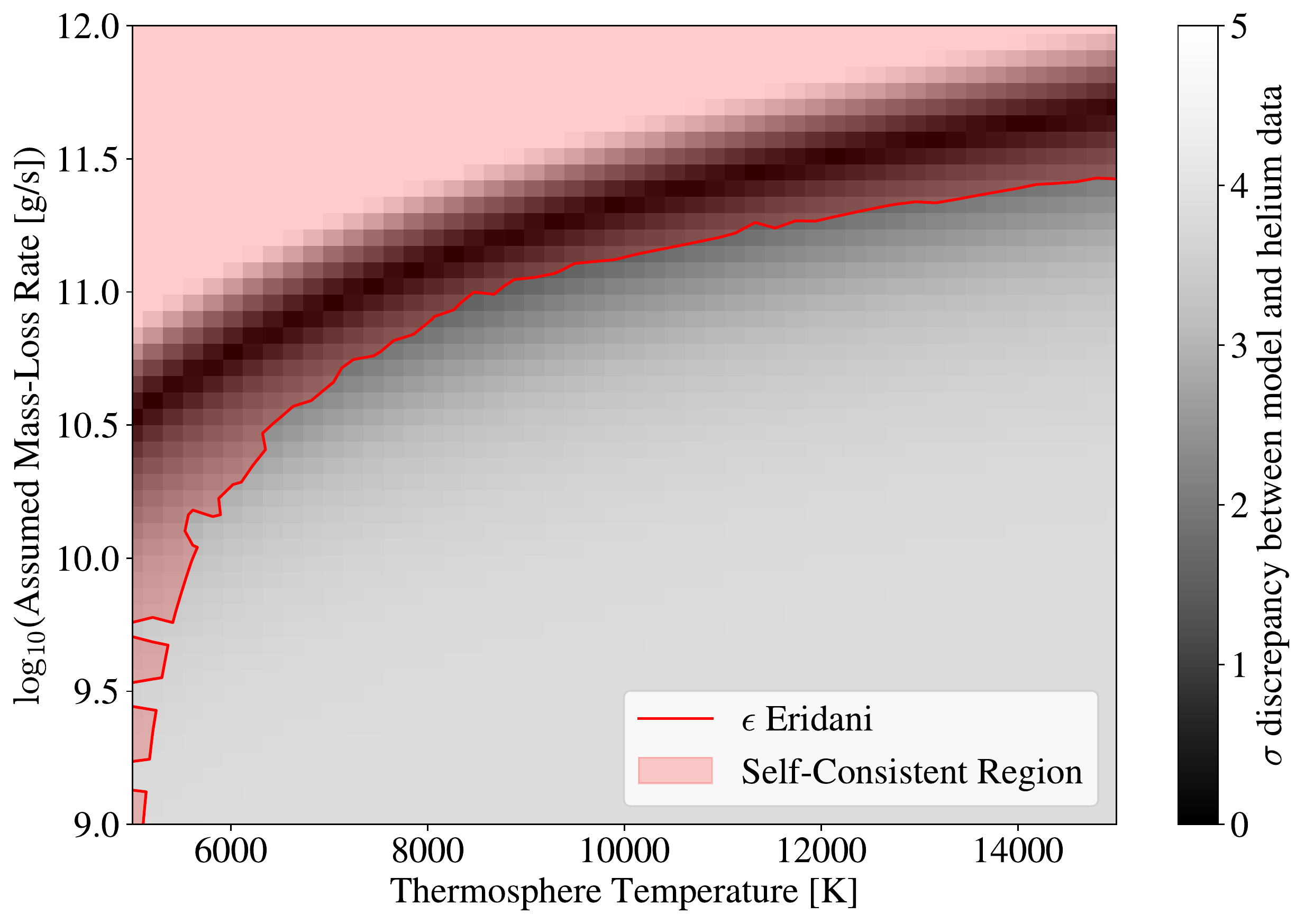}
\caption{Same as Figure~\ref{hp11}, but for the HAT-P-18b observations presented in \citet{Paragas21}.}
\label{hp18}
\end{figure}

HAT-P-18b is a Jupiter-sized, Saturn-mass planet ($\Phi_\mathrm{p} \approx 4 \times 10^{12}$~erg~g$^{-1}$) orbiting a K2 dwarf with a period of 5.5~days \citep{Hartman11}. In \cite{Paragas21}, we used narrowband photometry to detect helium absorption from an outflowing atmosphere with a significance of 4$\sigma$. In Figure~\ref{hp18}, we fit the measurement from that work using our Parker wind model and compare the resulting contour to the energetically self-consistent region. For the XUV spectrum, we used the MUSCLES spectrum of the young K2 dwarf $\epsilon$ Eridani. Though HAT-P-18b is rather old $12.4^{+4.4}_{-6.4}$~Gyr \citep{Hartman11}, it is quite active ($\log(R'_\mathrm{HK}) = -4.73$; H. Isaacson priv. comm.), so $\epsilon$ Eridani \citep[$\log(R'_\mathrm{HK}) = -4.51$;][]{Wright04} is a reasonable choice of XUV proxy from the MUSCLES catalog. 

For this planet, we could not limit the allowed mass-loss rate using energetics alone. The majority of models consistent with the helium data are also self-consistent. This is because the host star outputs a large XUV flux; as we discussed in Section~\ref{sec:hp11}, this tends to increase the number of self-consistent solutions with large mass-loss rates. More precise constraints on the mass-loss rate in this system will require precise line shape constraints from spectroscopic follow-up. The faintness (\textit{J} = 10.8) of this system makes spectroscopic follow-up with CARMENES difficult, as its \textit{J} magnitude is larger than the recommended magnitude limit for this instrument\footnote{https://carmenes.caha.es/ext/instrument/index.html}, so line-shape measurements will require time on larger facilities like Keck-II/NIRSPEC \citep[e.g.][]{Kirk20, Spake21}.

\section{Discussion and Conclusions} \label{sec:conc}
The isothermal Parker wind model is commonly used to interpret observations of metastable helium, with the ultimate goal of obtaining precise mass-loss rates for planets with observed helium absorption. However, there is a degeneracy between the outflow temperature and mass-loss rate in these models. The temperature can be constrained with spectroscopically-resolved observations at high SNR, but remains a large source of uncertainty when fitting unresolved measurements and/or those at a lower SNR. We partially resolved this degeneracy by determining the maximum mass-loss efficiency of an isothermal wind driven by photoionization. We found that the efficiency is limited by a quantity we call the heating coefficient, which is the product of the number density and a weighted photoionization cross section, as described by Equation~(\ref{finaleq}). We leveraged this constraint to show that a subset of the isothermal Parker wind models that agree with observations do not generate enough heat to remain energetically self-consistent. Outflows that generate just enough heat to remain self-consistent are exactly energy-limited, with the numerical value of the efficiency term depending on the assumed mass-loss rate, and outflows that generate excess heat must re-radiate some of their energy.

We re-examined published photometric and low-resolution spectroscopic observations of mass loss from HAT-P-11b, WASP-69b, and HAT-P-18b, which were originally fitted with a Parker wind model that allowed for a wide range of outflow temperatures and mass loss rates. We showed that the outflows from the former two planets must be relatively weak ($\dot{M} \lesssim 10^{11.5}$~g s$^{-1}$), but found that energetics could not further constrain the mass-loss rate for HAT-P-18b. Our results are in good agreement with more detailed numerical simulations of WASP-69b \citep{Wang21a} and HAT-P-11b \citep{Salz16a}, and additionally agree with complementary constraints on the outflow temperature of HAT-P-11b from line shape measurements \citep{Allart18, dosSantos21}. Furthermore, when line shape information is available (as it is for seven planets in the literature: HAT-P-11b, HAT-P-32b, WASP-69b, WASP-107b, HD 189733b, HD 209458b, and GJ 3470b) we can use our methodology to assess how close an outflow is to the energy limit.  In this study, we showed that HAT-P-11b is experiencing an energy-limited outflow.

Our investigation revealed that the location of the self-consistent region on the $\dot{M}-T_0$ plane can shift in response to changing assumptions. First, as we showed in Figure~\ref{hp11}, the extent of this region is sensitive to the assumed stellar spectrum; whenever possible, the spectra used in this model should be observationally calibrated. The models are also sensitive to the details of the photoionization calculation; any over- or under-estimation of the ionization fraction changes the self-consistent region. The assumed ratio of hydrogen to helium also matters, because the heating cross-section for neutral helium can be an order of magnitude larger than that for hydrogen for the late-type stars we considered in this work. Finally, the self-consistent region can move to higher temperatures if there are additional heat sources that are not included in the model. For instance, core-powered mass-loss \citep{Ginzburg16, Gupta19, Gupta20, Gupta21} could provide additional luminosity on the right-hand side of Equation~(\ref{proxy}).

In addition to refining the inferred mass-loss rates from metastable helium observations, our energetics framework also allows us to more generally elucidate the connection between the mass-loss efficiency in the energy limit and the outflow photophysics. This can be important when considering other potential heat sources. For example, \citet{Howe20} studied the potential heating from photodissociation of molecular hydrogen and found that it was a sub-dominant outflow driver when $\varepsilon$ was fixed to 0.1. Although our framework is imperfect for considering molecular winds, as there are more sources of heating and cooling to consider \citep{Glassgold15, Salz16a}, we can nonetheless use it to explore what happens when we relax the assumption of fixed efficiency. The cross-section for molecular hydrogen photodissociation is resonant at the frequencies of Lyman-Werner transitions \citep[e.g.][]{Heays17}, causing the gas to self-shield. This means that the optical depth to dissociating photons will quickly exceed unity in the outer region of the outflow, greatly reducing the heating rate per Equation~\ref{fullexpr} inside a thin photodissociated region \citep{Draine96}. We conclude that self-shielding may cause $\varepsilon$ to be even lower than assumed by \citet{Howe20}.

Population-level surveys of present-day mass loss are increasingly within reach. As the body of published helium absorption signals continues to grow, it is important to develop models that allow us to quickly and uniformly infer mass-loss rates from a large sample of observations. Our new outflow energetics framework can be used to enhance the scientific output of these surveys by providing more precise constraints on the retrieved mass-loss rates without significant computational expense. Although this method is currently limited to lower gravity planets, it could be extended to higher gravities in future studies by using simple prescriptions for the density structures and cooling rates of non-isothermal outflows. Despite this limitation, our framework can already be used to model helium signals from low-gravity, H/He-rich planets, like those in and near the Neptune desert. Helium observations have the potential to provide the first population-level constraints on the predicted mass-loss rates for these enigmatic planets.

\acknowledgments

We thank the anonymous referee for improving the quality of this paper and Antonija Oklop{\v{c}}i{\'c}, Yayaati Chachan, Konstantin Batygin, and Howard Isaacson for helpful conversations. SV is supported by an NSF Graduate Research Fellowship. HAK acknowledges support from NSF CAREER grant 1555095. LAdS acknowledges support from the European Research Council (ERC) under the European Union's Horizon 2020 research and innovation programme (grant agreement No 724427, project {\sc Four Aces}) and from the National Centre for Competence in Research PlanetS supported by the Swiss National Science Foundation (SNSF).

\facilities{ADS, NASA Exoplanet Archive}

\software{\texttt{numpy} \citep{Harris20},
\texttt{scipy} \citep{Virtanen20},
\texttt{astropy} \citep{Astropy13, Astropy18},
\texttt{matplotlib} \citep{Hunter07},
\texttt{p-winds} \citep{dosSantos21}}

\appendix 

\section{Non-Negligible Cooling}
\label{cooling}

In the recombination limit, cooling becomes important \citep{MurrayClay09, Owen16, Lampon21b}, and the approximation we made in Equation~(\ref{proxy}) greatly overpredicts the efficiency. We can repeat the calculation without making that assumption. When the cooling rate $\Lambda$ can be written exactly, the efficiency is:

\begin{align}
    \varepsilon_\mathrm{RL} = \frac{1}{R_\mathrm{p}^2} \int_{R_p}^{R_\mathrm{Roche}} r^2 \sum_i n_{i,r}\bar{\sigma}_{i,r} dr - \frac{4}{R_\mathrm{p}^2}\frac{1}{F_\mathrm{XUV}}\int_{R_\mathrm{p}}^{R_\mathrm{Roche}}\Lambda r^2 dr, \label{rr}
\end{align}

where the factor of four in the numerator of the cooling term reflects the fact that the planet can cool through all $4\pi$ steradians. Typically, Lyman-$\alpha$ is taken to be the dominant coolant for the outflow, in which case the cooling rate can be written:

\begin{equation}
\Lambda = C \Big(\frac{n_{\mathrm{H}^+, r}}{\mathrm{cm}^{-3}}\Big)\Big(\frac{n_{\mathrm{H}, r}}{\mathrm{cm}^{-3}}\Big)\exp{(-T_\mathrm{c}/T_0)},
\label{cool}
\end{equation}

where $n_{\mathrm{H}^+,r}$ is the number density profile of ionized hydrogen, $n_{\mathrm{H},r}$ is that of neutral hydrogen, $C = 7.5 \times 10^{-19}$~erg s$^{-1}$/cm$^3$, and $T_c = 118348$~K \citep{Black81, MurrayClay09, Owen16}. Rather than tracing out a permitted region on the $\dot{M}-T_0$ plane as in Equation~(\ref{finaleq}), Equation~(\ref{rr}) traces out a single curve of allowed solutions because the cooling is specified exactly. There are, however, many other cooling processes to consider even in a H/He gas including helium line emission, hydrogen and helium recombination cooling, and free-free emission \citep{Black81, Salz16a}. Additionally, our fundamental assumption of an isothermal wind is more readily violated for planets in the recombination limit when wind-launching is treated self-consistently \citep{Salz16a}. For these reasons, our framework is ill-suited to exactly treat planets with strong cooling, and we default to the upper-limit formulation in Equation~(\ref{consistency}).


\begin{thebibliography}{}
\bibitem[Adams(2011)]{Adams11} Adams, F.~C.\ 2011, \apj, 730, 27. doi:10.1088/0004-637X/730/1/27
\bibitem[Allart et al.(2018)]{Allart18} Allart, R., Bourrier, V., Lovis, C., et al.\ 2018, Science, 362, 1384. doi:10.1126/science.aat5879
\bibitem[Allart et al.(2019)]{Allart19} Allart, R., Bourrier, V., Lovis, C., et al.\ 2019, \aap, 623, A58. doi:10.1051/0004-6361/201834917
\bibitem[Anderson et al.(2014)]{Anderson14} Anderson, D.~R., Collier Cameron, A., Delrez, L., et al.\ 2014, \mnras, 445, 1114. doi:10.1093/mnras/stu1737
\bibitem[Astropy Collaboration et al.(2013)]{Astropy13} Astropy Collaboration, Robitaille, T.~P., Tollerud, E.~J., et al.\ 2013, \aap, 558, A33. doi:10.1051/0004-6361/201322068
\bibitem[Astropy Collaboration et al.(2018)]{Astropy18} Astropy Collaboration, Price-Whelan, A.~M., Sip{\H{o}}cz, B.~M., et al.\ 2018, \aj, 156, 123. doi:10.3847/1538-3881/aabc4f
\bibitem[Bakos et al.(2010)]{Bakos10} Bakos, G. {\'A}., Torres, G., P{\'a}l, A., et al.\ 2010, \apj, 710, 1724. doi:10.1088/0004-637X/710/2/1724
\bibitem[Black(1981)]{Black81} Black, J.~H.\ 1981, \mnras, 197, 553. doi:10.1093/mnras/197.3.553
\bibitem[Carolan et al.(2020)]{Carolan20} Carolan, S., Vidotto, A.~A., Plavchan, P., et al.\ 2020, \mnras, 498, L53. doi:10.1093/mnrasl/slaa127
\bibitem[dos Santos et al.(2021)]{dosSantos21} dos Santos, L.~A., Vidotto, A.~A., Vissapragada, S., et al.\ 2021, arXiv:2111.11370
\bibitem[Draine \& Bertoldi(1996)]{Draine96} Draine, B.~T. \& Bertoldi, F.\ 1996, \apj, 468, 269. doi:10.1086/177689
\bibitem[Erkaev et al.(2007)]{Erkaev07} Erkaev, N.~V., Kulikov, Y.~N., Lammer, H., et al.\ 2007, \aap, 472, 329. doi:10.1051/0004-6361:20066929
\bibitem[France et al.(2016)]{France16} France, K., Loyd, R.~O.~P., Youngblood, A., et al.\ 2016, \apj, 820, 89. doi:10.3847/0004-637X/820/2/89
\bibitem[Fulton et al.(2017)]{Fulton17} Fulton, B.~J., Petigura, E.~A., Howard, A.~W., et al.\ 2017, \aj, 154, 109.
doi:10.3847/1538-3881/aa80eb
\bibitem[Fulton \& Petigura(2018)]{Fulton18} Fulton, B.~J. \& Petigura, E.~A.\ 2018, \aj, 156, 264. doi:10.3847/1538-3881/aae828
\bibitem[Gaidos et al.(2020a)]{Gaidos20a} Gaidos, E., Hirano, T., Mann, A.~W., et al.\ 2020, \mnras, 495, 650. doi:10.1093/mnras/staa918
\bibitem[Gaidos et al.(2020b)]{Gaidos20b} Gaidos, E., Hirano, T., Wilson, D.~J., et al.\ 2020, \mnras, 498, L119. doi:10.1093/mnrasl/slaa136
\bibitem[Ginzburg et al.(2016)]{Ginzburg16} Ginzburg, S., Schlichting, H.~E., \& Sari, R.\ 2016, \apj, 825, 29. doi:10.3847/0004-637X/825/1/29
\bibitem[Glassgold \& Najita(2015)]{Glassgold15} Glassgold, A.~E. \& Najita, J.~R.\ 2015, \apj, 810, 125. doi:10.1088/0004-637X/810/2/125
\bibitem[Guo \& Ben-Jaffel(2016)]{Guo16} Guo, J.~H. \& Ben-Jaffel, L.\ 2016, \apj, 818, 107. doi:10.3847/0004-637X/818/2/107
\bibitem[Gupta \& Schlichting(2019)]{Gupta19} Gupta, A. \& Schlichting, H.~E.\ 2019, \mnras, 487, 24. doi:10.1093/mnras/stz1230
\bibitem[Gupta \& Schlichting(2020)]{Gupta20} Gupta, A. \& Schlichting, H.~E.\ 2020, \mnras, 493, 792. doi:10.1093/mnras/staa315
\bibitem[Gupta \& Schlichting(2021)]{Gupta21} Gupta, A. \& Schlichting, H.~E.\ 2021, \mnras, 504, 4634. doi:10.1093/mnras/stab1128
\bibitem[Hardegree-Ullman et al.(2020)]{HardegreeUllman20} Hardegree-Ullman, K.~K., Zink, J.~K., Christiansen, J.~L., et al.\ 2020, \apjs, 247, 28. doi:10.3847/1538-4365/ab7230
\bibitem[Harris et al.(2020)]{Harris20} Harris, C.~R., Millman, K.~J., van der Walt, S.~J., et al.\ 2020, \nat, 585, 357. doi:10.1038/s41586-020-2649-2
\bibitem[Hartman et al.(2011)]{Hartman11} Hartman, J.~D., Bakos, G. {\'A}., Sato, B., et al.\ 2011, \apj, 726, 52. doi:10.1088/0004-637X/726/1/52
\bibitem[Heays et al.(2017)]{Heays17} Heays, A.~N., Bosman, A.~D., \& van Dishoeck, E.~F.\ 2017, \aap, 602, A105. doi:10.1051/0004-6361/201628742
\bibitem[Hirano et al.(2020)]{Hirano20} Hirano, T., Krishnamurthy, V., Gaidos, E., et al.\ 2020, \apjl, 899, L13. doi:10.3847/2041-8213/aba6eb
\bibitem[Howe et al.(2020)]{Howe20} Howe, A.~R., Adams, F.~C., \& Meyer, M.~R.\ 2020, \apj, 894, 130. doi:10.3847/1538-4357/ab620c
\bibitem[Hunter(2007)]{Hunter07} Hunter, J.~D.\ 2007, Computing in Science and Engineering, 9, 90. doi:10.1109/MCSE.2007.55
\bibitem[Kasper et al.(2020)]{Kasper20} Kasper, D., Bean, J.~L., Oklop{\v{c}}i{\'c}, A., et al.\ 2020, \aj, 160, 258. doi:10.3847/1538-3881/abbee6
\bibitem[Kirk et al.(2020)]{Kirk20} Kirk, J., Alam, M.~K., L{\'o}pez-Morales, M., et al.\ 2020, \aj, 159, 115. doi:10.3847/1538-3881/ab6e66
\bibitem[Krenn et al.(2021)]{Krenn21} Krenn, A.~F., Fossati, L., Kubyshkina, D., et al.\ 2021, \aap, 650, A94. doi:10.1051/0004-6361/202140437
\bibitem[Krishnamurthy et al.(2021)]{Krishnamurthy21} Krishnamurthy, V., Hirano, T., Stef{\'a}nsson, G., et al.\ 2021, \aj, 162, 82. doi:10.3847/1538-3881/ac0d57
\bibitem[Kubyshkina et al.(2018)]{Kubyshkina18} Kubyshkina, D., Fossati, L., Erkaev, N.~V., et al.\ 2018, \apjl, 866, L18. doi:10.3847/2041-8213/aae586
\bibitem[Lamers \& Cassinelli(1999)]{Lamers99} Lamers, H.~J.~G.~L.~M. \& Cassinelli, J.~P.\ 1999, Introduction to Stellar Winds, by Henny J. G. L. M. Lamers and Joseph P. Cassinelli, pp. 452. ISBN 0521593980. Cambridge, UK: Cambridge University Press, June 1999., 452
\bibitem[Lamp{\'o}n et al.(2020)]{Lampon20} Lamp{\'o}n, M., L{\'o}pez-Puertas, M., Lara, L.~M., et al.\ 2020, \aap, 636, A13. doi:10.1051/0004-6361/201937175
\bibitem[Lamp{\'o}n et al.(2021a)]{Lampon21a} Lamp{\'o}n, M., L{\'o}pez-Puertas, M., Sanz-Forcada, J., et al.\ 2021, \aap, 647, A129. doi:10.1051/0004-6361/202039417
\bibitem[Lamp{\'o}n et al.(2021b)]{Lampon21b} Lamp{\'o}n, M., L{\'o}pez-Puertas, M., Czesla, S., et al.\ 2021, \aap, 648, L7. doi:10.1051/0004-6361/202140423
\bibitem[Lopez \& Fortney(2013)]{Lopez13} Lopez, E.~D., \& Fortney, J.~J.\ 2013, \apj, 776, 2 
\bibitem[Loyd et al.(2016)]{Loyd16} Loyd, R.~O.~P., France, K., Youngblood, A., et al.\ 2016, \apj, 824, 102. doi:10.3847/0004-637X/824/2/102
\bibitem[MacLeod \& Oklop{\v{c}}i{\'c}(2021)]{MacLeod21} MacLeod, M. \& Oklop{\v{c}}i{\'c}, A.\ 2021, arXiv:2107.07534
\bibitem[Mansfield et al.(2018)]{Mansfield18} Mansfield, M., Bean, J.~L., Oklop{\v{c}}i{\'c}, A., et al.\ 2018, \apjl, 868, L34
\bibitem[McCann et al.(2019)]{McCann19} McCann, J., Murray-Clay, R.~A., Kratter, K., et al.\ 2019, \apj, 873, 89. doi:10.3847/1538-4357/ab05b8
\bibitem[Murray-Clay et al.(2009)]{MurrayClay09} Murray-Clay, R.~A., Chiang, E.~I., \& Murray, N.\ 2009, \apj, 693, 23. doi:10.1088/0004-637X/693/1/23
\bibitem[Ninan et al.(2020)]{Ninan20} Ninan, J.~P., Stefansson, G., Mahadevan, S., et al.\ 2020, \apj, 894, 97. doi:10.3847/1538-4357/ab8559
\bibitem[Nortmann et al.(2018)]{Nortmann18} Nortmann, L., Pall{\'e}, E., Salz, M., et al.\ 2018, Science, 362, 1388
\bibitem[Oklop{\v{c}}i{\'c} \& Hirata(2018)]{Oklopcic18} Oklop{\v{c}}i{\'c}, A. \& Hirata, C.~M.\ 2018, \apjl, 855, L11. doi:10.3847/2041-8213/aaada9
\bibitem[Oklop{\v{c}}i{\'c}(2019)]{Oklopcic19} Oklop{\v{c}}i{\'c}, A.\ 2019, \apj, 881, 133. doi:10.3847/1538-4357/ab2f7f
\bibitem[Osterbrock \& Ferland(2006)]{Osterbrock06} Osterbrock, D.~E. \& Ferland, G.~J.\ 2006, Astrophysics of gaseous nebulae and active galactic nuclei, 2nd. ed. by D.E. Osterbrock and G.J. Ferland. Sausalito, CA: University Science Books, 2006
\bibitem[Owen(2019)]{Owen19} Owen, J.~E.\ 2019, Annual Review of Earth and Planetary Sciences, 47, 67
\bibitem[Owen \& Adams(2014)]{Owen14} Owen, J.~E. \& Adams, F.~C.\ 2014, \mnras, 444, 3761. doi:10.1093/mnras/stu1684
\bibitem[Owen \& Alvarez(2016)]{Owen16} Owen, J.~E. \& Alvarez, M.~A.\ 2016, \apj, 816, 34. doi:10.3847/0004-637X/816/1/34
\bibitem[Owen \& Jackson(2012)]{Owen12} Owen, J.~E. \& Jackson, A.~P.\ 2012, \mnras, 425, 2931. doi:10.1111/j.1365-2966.2012.21481.x
\bibitem[Owen \& Wu(2013)]{Owen13} Owen, J.~E. \& Wu, Y.\ 2013, \apj, 775, 105
\bibitem[Owen \& Wu(2017)]{Owen17} Owen, J.~E. \& Wu, Y.\ 2017, \apj, 847, 29
\bibitem[Paragas et al.(2021)]{Paragas21} Paragas, K., Vissapragada, S., Knutson, H.~A., et al.\ 2021, \apjl, 909, L10. doi:10.3847/2041-8213/abe706
\bibitem[Salz et al.(2016a)]{Salz16a} Salz, M., Schneider, P.~C., Czesla, S., et al.\ 2016, \aap, 585, L2. doi:10.1051/0004-6361/201527042
\bibitem[Salz et al.(2016b)]{Salz16b} Salz, M., Czesla, S., Schneider, P.~C., et al.\ 2016, \aap, 586, A75. doi:10.1051/0004-6361/201526109
\bibitem[Salz et al.(2018)]{Salz18} Salz, M., Czesla, S., Schneider, P.~C., et al.\ 2018, \aap, 620, A97
\bibitem[Seager(2010)]{Seager10} Seager, S.\ 2010, Exoplanet Atmospheres: Physical Processes.  By Sara Seager.  Princeton University Press, 2010. ISBN: 978-1-4008-3530-0
\bibitem[Seidel et al.(2021)]{Seidel21} Seidel, J.~V., Ehrenreich, D., Allart, R., et al.\ 2021, \aap, 653, A73. doi:10.1051/0004-6361/202140569
\bibitem[Shematovich et al.(2014)]{Shematovich14} Shematovich, V.~I., Ionov, D.~E., \& Lammer, H.\ 2014, \aap, 571, A94. doi:10.1051/0004-6361/201423573
\bibitem[Spake et al.(2018)]{Spake18} Spake, J.~J., Sing, D.~K., Evans, T.~M., et al.\ 2018, \nat, 557, 68. doi:10.1038/s41586-018-0067-5
\bibitem[Spake et al.(2021)]{Spake21} Spake, J.~J., Oklop{\v{c}}i{\'c}, A., \& Hillenbrand, L.~A.\ 2021, \aj, 162, 284. doi:10.3847/1538-3881/ac178a
\bibitem[Stone \& Proga(2009)]{Stone09} Stone, J.~M. \& Proga, D.\ 2009, \apj, 694, 205. doi:10.1088/0004-637X/694/1/205
\bibitem[Trammell et al.(2011)]{Trammell11} Trammell, G.~B., Arras, P., \& Li, Z.-Y.\ 2011, \apj, 728, 152. doi:10.1088/0004-637X/728/2/152
\bibitem[Trammell et al.(2014)]{Trammell14} Trammell, G.~B., Li, Z.-Y., \& Arras, P.\ 2014, \apj, 788, 161. doi:10.1088/0004-637X/788/2/161
\bibitem[Van Eylen et al.(2018)]{vanEylen18} Van Eylen, V., Agentoft, C., Lundkvist, M.~S., et al.\ 2018, \mnras, 479, 4786. doi:10.1093/mnras/sty1783
\bibitem[Vidal-Madjar et al.(2003)]{vidalMadjar03} Vidal-Madjar, A., Lecavelier des Etangs, A., D{\'e}sert, J.-M., et al.\ 2003, \nat, 422, 143. doi:10.1038/nature01448
\bibitem[Vidal-Madjar et al.(2004)]{vidalMadjar04} Vidal-Madjar, A., D{\'e}sert, J.-M., Lecavelier des Etangs, A., et al.\ 2004, \apjl, 604, L69. doi:10.1086/383347
\bibitem[Virtanen et al.(2020)]{Virtanen20} Virtanen, P., Gommers, R., Oliphant, T.~E., et al.\ 2020, Nature Methods, 17, 261. doi:10.1038/s41592-019-0686-2
\bibitem[Vissapragada et al.(2020)]{Vissapragada20} Vissapragada, S., Knutson, H.~A., Jovanovic, N., et al.\ 2020, \aj, 159, 278. doi:10.3847/1538-3881/ab8e34
\bibitem[Wang \& Dai(2021a)]{Wang21a} Wang, L. \& Dai, F.\ 2021, \apj, 914, 98. doi:10.3847/1538-4357/abf1ee
\bibitem[Wang \& Dai(2021b)]{Wang21b} Wang, L. \& Dai, F.\ 2021, \apj, 914, 99. doi:10.3847/1538-4357/abf1ed
\bibitem[Watson et al.(1981)]{Watson81} Watson, A.~J., Donahue, T.~M., \& Walker, J.~C.~G.\ 1981, \icarus, 48, 150. doi:10.1016/0019-1035(81)90101-9
\bibitem[Wright et al.(2004)]{Wright04} Wright, J.~T., Marcy, G.~W., Butler, R.~P., et al.\ 2004, \apjs, 152, 261. doi:10.1086/386283
\bibitem[Yan \& Henning(2018)]{Yan18} Yan, F. \& Henning, T.\ 2018, Nature Astronomy, 2, 714. doi:10.1038/s41550-018-0503-3
\bibitem[Yan et al.(1998)]{Yan98} Yan, M., Sadeghpour, H.~R., \& Dalgarno, A.\ 1998, \apj, 496, 1044. doi:10.1086/305420
\bibitem[Youngblood et al.(2016)]{Youngblood16} Youngblood, A., France, K., Loyd, R.~O.~P., et al.\ 2016, \apj, 824, 101. doi:10.3847/0004-637X/824/2/101
\bibitem[Zhang et al.(2021)]{Zhang21} Zhang, M., Knutson, H.~A., Wang, L., et al.\ 2021, \aj, 161, 181. doi:10.3847/1538-3881/abe382
\end{thebibliography}
\end{document}